\documentclass[aps,pra,twocolumn reprint, notitlepage, showpacs, floatfix, superscriptaddress, longbibliography]{revtex4-1}

\usepackage{silence}
\usepackage{amsthm,amsmath,amsfonts,amssymb,verbatim, color}
\usepackage{graphicx}
\usepackage[percent]{overpic}
\usepackage{epsfig,slashed}
\usepackage{epstopdf}
\usepackage{lipsum}
\usepackage{float}
\usepackage{mathtools}
\usepackage[colorlinks=true,citecolor=blue,linkcolor=black,urlcolor=blue]{hyperref}
\usepackage{natbib}
\usepackage{subfigure}
\usepackage[mathscr]{euscript}
\usepackage[export]{adjustbox}
\usepackage[bottom]{footmisc}
\usepackage{textcomp}
\usepackage[margin=1.25in]{geometry}
\usepackage{wrapfig}

\makeatletter
\newsavebox\myboxA
\newsavebox\myboxB
\newlength\mylenA
\newcommand*\xoverline[2][0.75]{%
    \sbox{\myboxA}{$\m@th#2$}%
    \setbox\myboxB\null
    \ht\myboxB=\ht\myboxA%
    \dp\myboxB=\dp\myboxA%
    \wd\myboxB=#1\wd\myboxA
    \sbox\myboxB{$\m@th\overline{\copy\myboxB}$}
    \setlength\mylenA{\the\wd\myboxA}
    \addtolength\mylenA{-\the\wd\myboxB}%
    \ifdim\wd\myboxB<\wd\myboxA%
       \rlap{\hskip 0.5\mylenA\usebox\myboxB}{\usebox\myboxA}%
    \else
        \hskip -0.5\mylenA\rlap{\usebox\myboxA}{\hskip 0.5\mylenA\usebox\myboxB}%
    \fi}
\makeatother

\begin{document}

\title{Localization in one-dimensional relativistic quantum mechanics}

\author{Abhay Mehta}
\affiliation{St. Xavier's College, Mumbai 400001, India}

\author{Sandeep Joshi\footnote{sjoshi@barc.gov.in}}
\affiliation{Nuclear Physics Division, Bhabha Atomic Research Centre, Mumbai 400085, India}
\affiliation{Homi Bhabha National Institute, Training School Complex, Anushakti Nagar, Mumbai 400094, India}

\author{Sudhir R. Jain\footnote{srjain@barc.gov.in}}
\affiliation{Nuclear Physics Division, Bhabha Atomic Research Centre, Mumbai 400085, India}
\affiliation{Homi Bhabha National Institute, Training School Complex, Anushakti Nagar, Mumbai 400094, India}
\affiliation{UM-DAE Centre for Excellence in Basic Sciences, University of Mumbai, Vidynagari Campus, Mumbai 400098, India}

\begin{abstract}
We present the relativistic analogue of Anderson localization in one dimension. We use Dirac equation to calculate the transmission probability for a spin-$\frac{1}{2}$ particle incident upon a rectangular barrier. Using the transfer matrix formalism, we numerically compute the transmission probability for the case of a large number of identical barriers spread randomly in one dimension. The particular case when the incident particle has three component momentum and shows spin-flip phenomena is also considered. Our calculations suggest that the incident relativistic particle shows localization behaviour similar to that of Anderson localization. A number of results which are generalizations of the non-relativistic case are also obtained. 
\end{abstract}

\maketitle
\section{Introduction}
Anderson localization \cite{Anderson} has occupied centrepiece in the development of condensed matter physics. The beautiful state of the art argument by Anderson has been simplified a long time ago by Mott and Twose \cite{mott}. In addition, Anderson localization appears in quantum chaos as dynamical localization in kicked rotor \cite{prange} and Fermi-Ulam model \cite{jain1,jain2}. This phenomenon is also experimentally demonstrated in the physics of cold atoms \cite{schleich,raizen}. In the context of disordered solids, Anderson's work prompted many papers and experiments on localization \cite{Eg_1}\cite{Eg_2}\cite{Eg_1_1}\cite{Eg_1_2}.

Mott and Twose simplified the Anderson's model by considering an infinite array of rectangular potential barriers with some disorder incorporated through the positions of the barriers or via their heights. Anderson had introduced disorder in the structure by arbitrarily varying the energy at each lattice site.
Here, we consider the relativistic equivalent of Anderson model and consider the solutions of the Dirac equation. We first calculate the transmission probability of an incident spin one-half particle over a finite rectangular barrier and subsequently, compute the same for a number of identical barriers arranged randomly in space. Relativistic quantum mechanics presents some situations which have no analogue in non-relativistic quantum mechanics. For instance, for a one-dimensional arrangement of barriers, if the momentum of the incident particle is taken with one component along the arrangement of barriers, the spin of the particle does not flip. On the other hand, if all momentum components are considered, then the transmitted particles would include those also whose spin is flipped \cite{Glass}. 

Moreover, in the relativistic case, the tunneling for energy below the top of the barrier will be accompanied by the complications arising due to pair production and Klein paradox \cite{Sakurai,BjorkenDrell}. In  this work, we do not address these situations. Thus, the energies considered are above the barrier.   

The layout of the paper is as follows: Section 2 defines and introduces the problem, Section 3 presents the formalism and calculations for a single barrier followed by results on localization. Section 4 repeats the methodology followed in the former section but for the case of spin-flip. Section 5 concludes with the main results of the paper.

\section{Dirac particle over a rectangular barrier}
Consider a spin-$\frac{1}{2}$ particle, propagating along the $z$ axis, incident upon a rectangular potential $V_{0}$ of width $a$. The motion of the particle is described by the Dirac equation \cite{Sakurai} :
\begin{equation}
\left(\gamma_{\mu} \frac{\partial}{\partial x_\mu}+ \frac{mc}{\hbar}\right) \psi = 0,
\end{equation}
where $\psi$ is a \textit{four}-component wave-function and $\gamma_{\mu}$, with $\mu = 1, 2, 3, 4$, are $4 \times 4$ Dirac matrices,  given by:
\begin{equation}
\gamma_{k} = 
\begin{pmatrix}
		0 & -i\sigma_{k}\\
        i\sigma_{k} & 0\\
\end{pmatrix} \quad \text{and}  \quad 
\gamma_{4} = 
\begin{pmatrix}
		I & 0\\
        0 & I\\
\end{pmatrix},
\end{equation}
where $\sigma_k$ are the three Pauli matrices and $I$ is a $2 \times 2$ identity matrix. 

A rectangular potential of width $a$ is defined as 
\begin{equation}
V(z) = 
\left\{
	\begin{array}{ll}
		0, \qquad   z < 0, \\
        V_{0},\quad \ 0 < z < a, \\
		0, \qquad  z > a
	\end{array}
\right.
\end{equation}

\begin{figure}[h] 
\centering
\includegraphics[width=\textwidth]{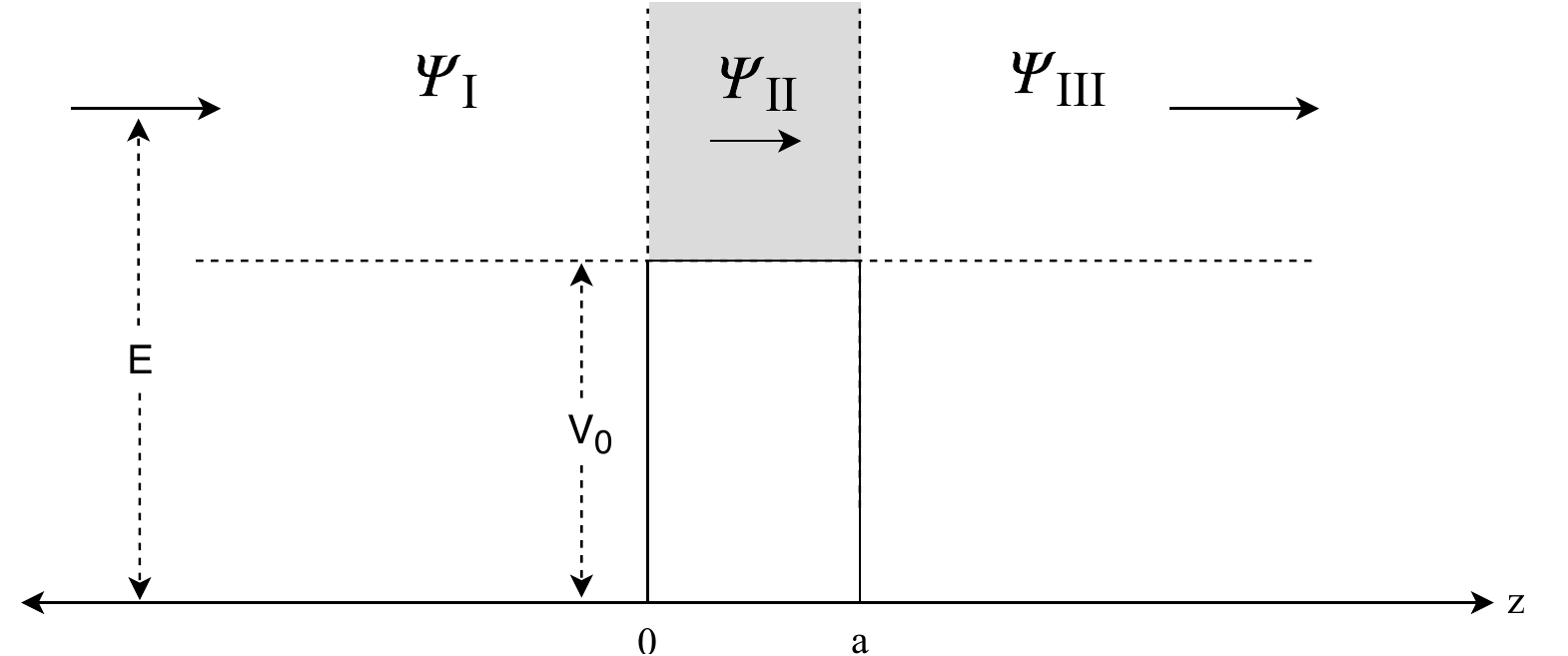}
\caption{Pictorial representation of a Dirac particle incident on a potential barrier with $E>V_{0}$. The darkened area represents the area over which the particle is above the barrier.}
\label{fig:1}
\end{figure}

In order to avoid the realm of the Klein Paradox \cite{Sakurai,BjorkenDrell}, we take the energy $E$ of the incident particle, such that $E > V_0 - mc^2$. We employ the transfer matrix approach to find the transmission and reflection coefficients. We consider two cases for momentum of the incident particle :$(a)$ particle carrying the momentum only along the incident direction and $(b)$ momentum along all three directions. For the second case the particle can undergo spin-flip upon incident on the barrier \cite{Glass}.
\section{One-component momentum}
\subsection{Transfer matrix formalism}
The general positive energy solutions of the Dirac equation in the three different regions in Fig.  \ref{fig:1} can be expressed as a superposition of plane waves:
\begin{alignat}{1}
\Psi_{I} (z) &= Au_{1}(p) e^{\frac{ip_{z}z}{\hbar}} + Bu_{2}(p) e^{\frac{ip_{z}z}{\hbar}} + Cu_{1}(-p)e^{\frac{-ip_{z}z}{\hbar}} + Du_{2}(-p)e^{\frac{-ip_{z}z}{\hbar}}, \nonumber 
\\
\Psi_{II} (z) &= Eu_{1}(q) e^{\frac{iq_{z}z}{\hbar}} + Fu_{2}(q) e^{\frac{iq_{z}z}{\hbar}} + Gu_{1}(-q)e^{\frac{-iq_{z}z}{\hbar}} + Hu_{2}(-q)e^{\frac{-iq_{z}z}{\hbar}},  
\\
\Psi_{III} (z) &=  Pu_{1}(p) e^{\frac{ip_{z}z}{\hbar}} + Qu_{2}(p) e^{\frac{ip_{z}z}{\hbar}} + Ru_{1}(-p)e^{\frac{-ip_{z}z}{\hbar}} + 	Su_{2}(-p)e^{\frac{-ip_{z}z}{\hbar}}, \nonumber
\end{alignat}
where $u_{1}$ and $u_{2}$ are the usual positive energy spin-up and spin-down spinors \cite{Sakurai}. Here $p_{z}$ is the momentum of the particle in free space and $q_{z}$ is the momentum of the particle over the barrier. Let us first consider the case when the particle carry only the momentum component in incident direction. For brevity, we shall henceforth write $p_{z}$ and $q_{z}$ as $p$ and $q$ respectively. It is well-known that there is no spin flip \cite{BjorkenDrell} in this case. This allows us to simplify (4) by ignoring the degeneracies in spin. Eq. (4) is then rewritten as :
\begin{alignat}{1}
		\Psi_{I} (z) = Au_{1}(p) e^{\frac{ipz}{\hbar}} + Cu_{1}(-p)e^{\frac{-ipz}{\hbar}}, \nonumber \\
		\Psi_{II} (z) = Eu_{1}(q) e^{\frac{iqz}{\hbar}} + Gu_{1}(-q)e^{\frac{-iqz}{\hbar}}, \\
		\Psi_{III} (z) =  Pu_{1}(p) e^{\frac{ipz}{\hbar}} + Ru_{1}(-p)e^{\frac{-ipz}{\hbar}}.\nonumber
\end{alignat}
According to the definition of the transfer matrix \cite{Markos}, the wave-function on both sides of the potential are connected by the equation:
\begin{equation}
\begin{pmatrix}
		P  e^{\frac{ipa}{\hbar}} \\
        R e^{\frac{-ipa}{\hbar}} \\
\end{pmatrix}  = M
\begin{pmatrix}
		A \\
        C \\
\end{pmatrix}.
\end{equation}
It is now possible to solve for the continuity of the Dirac equation at $z = 0$ and $z = a$ and obtain the transfer matrix $M$. We can, however,  simplify the calculation by  calculating the transfer matrix $M_{step}$ of the potential step at $z = 0$. The transfer matrix $M_{step}$ is given by:
\begin{equation}
\begin{pmatrix}
		E \\
        G \\
\end{pmatrix}  = M_{step}
\begin{pmatrix}
		A \\
        C \\
\end{pmatrix}.
\end{equation}
We solve (5) by evaluating the spinor $u_{1}$ and obtain the matrix equation:

\begin{equation}
\begin{pmatrix}
		1 & 1 \\
        p & -p \\
\end{pmatrix} 
\begin{pmatrix}
		A \\
        C \\
\end{pmatrix}  =
\begin{pmatrix}
		1 & 1 \\
        r q & -r q \\
\end{pmatrix} 
\begin{pmatrix}
		E \\
        G \\
\end{pmatrix}.
\end{equation}\\
where
\begin{equation}
r = \frac{E + mc^2}{E - V_{0} + mc^2} = \frac{1}{1 - \frac{V_{0}}{E + mc^2}}
\end{equation}
Solving (8) yields  $M_{step}$ as:
\begin{equation}
 M_{step} =
\begin{pmatrix}
\frac{1}{2} + \frac{p}{2rq} & \frac{1}{2} - \frac{p}{2rq} \\
\frac{1}{2} - \frac{p}{2rq} & \frac{1}{2} + \frac{p}{2rq} \\
\end{pmatrix}
\end{equation}
The transfer matrix $M$ for the rectangular potential can now be found with :
\begin{equation}
M = M_{step}^{-1}.M_{0}.M_{step},
\end{equation}
where $M_{0}$ is the transfer matrix of propagation inside a barrier of length $a$,
\begin{equation}
M_{0} = 
\begin{pmatrix}
		e^{\frac{iqa}{\hbar}} & 0\\
        0 & e^{\frac{-iqa}{\hbar}} \\
\end{pmatrix}.
\end{equation}\\
From (11), the transfer matrix M is given by : 
\begin{equation}
M = 
\begin{pmatrix}
		u & v\\
        v^* & u^* \\
\end{pmatrix};
\end{equation}
the elements $u$ and $v$ being
\begin{alignat}{1}
u &= \cos \left(\frac{aq}{\hbar}\right) + i\alpha_{+}\sin \left(\frac{aq}{\hbar}\right)\\
v &= +i\alpha_{-}\sin \left(\frac{aq}{\hbar}\right) \nonumber
\end{alignat}
where 
\begin{equation}
\alpha_{\pm} = \frac{1}{2}\left(\frac{p}{rq} \pm \frac{rq}{p}\right). 
\end{equation}
The transfer matrix $M$ maintains both time reversal symmetry and conservation of current density. This is verified by the calculation of the following relations \cite{Markos}:
\begin{alignat}{1}
\mbox{Det } M &= 1 \quad \mbox{(time-reversal symmetry),}\\
M^\dagger 
\begin{pmatrix}
		1 & 0\\
        0 & -1 \\
\end{pmatrix} M &= \begin{pmatrix}
		1 & 0\\
        0 & -1 \\
\end{pmatrix} \quad \mbox{(conservation of current density).}
\end{alignat}
\subsection{Transmission over one barrier}
\begin{figure}[h!]
\centering
\includegraphics[width=.75\textwidth]{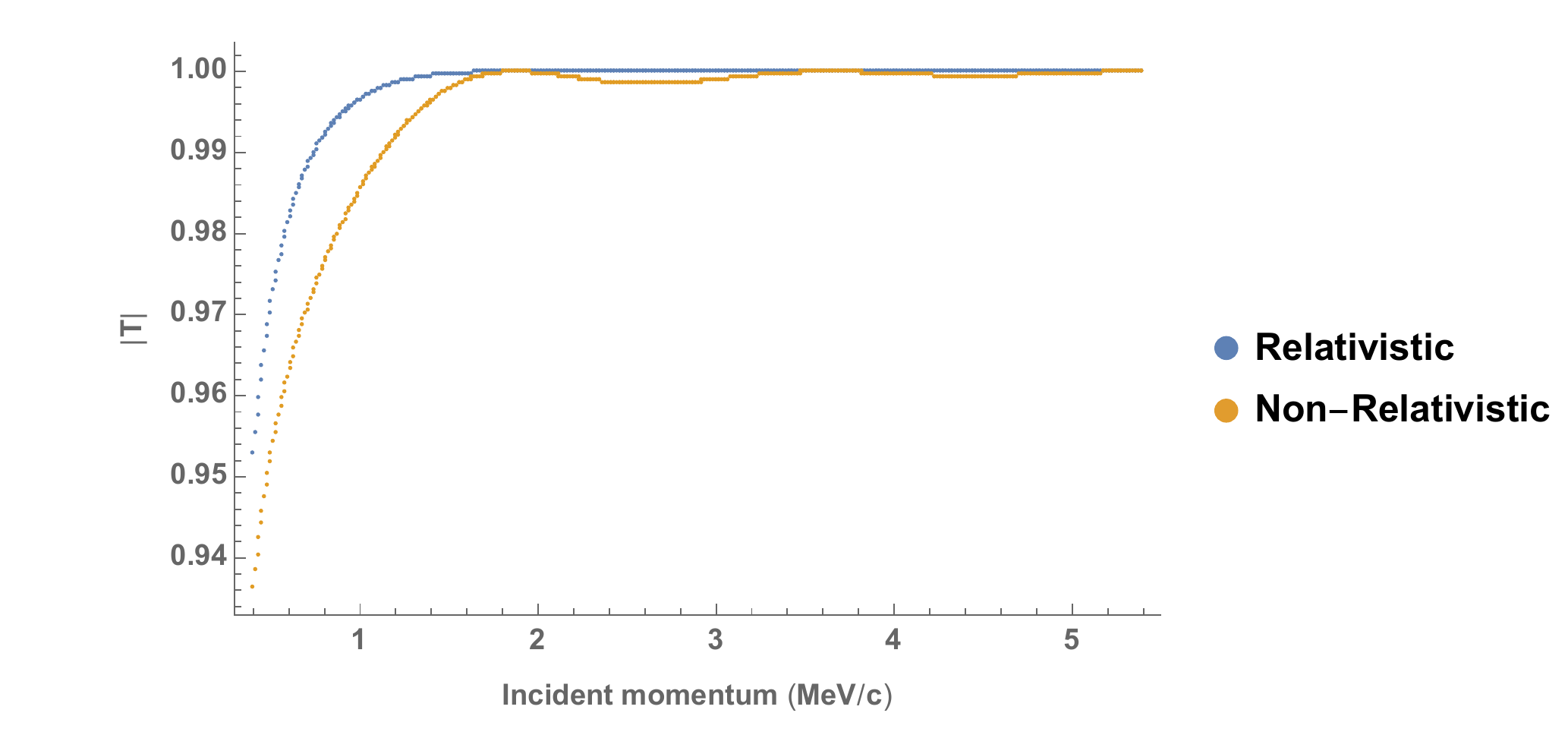}
\caption{The plot shows the transmission probability of an electron over one barrier for increasing values of incident momenta 'p' of the particle calculated separately using the relativistic and non-relativistic formula. Here the height and width of the barrier are 0.10 MeV and 350 fm respectively. The transmission in the relativistic case is minimum when energy of the incident particle is only slightly greater than barrier potential V.
}
\label{fig:2}
\end{figure}

Due to the degeneracy in spin, the transfer matrix for the relativistic case has the same formulation as that of the non-relativistic one. The transmission coefficient $T$ for a particle transmitted through a single rectangular barrier \cite{Markos} can simply be obtained from the transfer matrix $M$ as:
\begin{equation}
T = |t|^2 = \frac{1}{|M_{22}|^2} 
\end{equation}
Using the expression for $M_{22}$ from (13) we obtain $T$ as:
\begin{equation}
T =\bigg [ 1 + \frac{1}{4} \bigg (\frac{p}{rq} - \frac{rq}{p}\bigg )^2
	\text{sin}^2 \bigg  [\frac{aq}{\hbar}\bigg ] \bigg]^{-1}
\end{equation}
where the initial wave-functions are described in (5). Equation (19) reduces to the non-relativistic limit obtained from Schrodinger's equation under two conditions. First, if $r \rightarrow 1$ which implies that either the height of the potential barrier goes to 0, i.e. $V \rightarrow 0$ or the incident energy of the particle is much larger than the height of the potential barrier, i.e. $E \gg V$, and second if $r \rightarrow \pm \frac{p^2}{q^2}$. The latter being the ratio of kinetic energy in two regions in the non-relativistic limit $E \approx p^2/2m$. We  compare  results  given  by  equation  (19) to  its non-relativistic counterpart for dependency on the incident momentum of the particle in Fig. \ref{fig:2}.

\newpage
\subsection{Localization in one dimension}
\begin{figure}[ht]
    \centering
    \includegraphics[width=\textwidth]{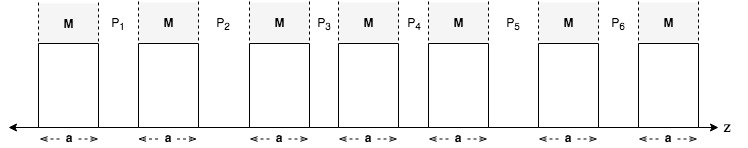}
    \caption{A finite portion of an infinite array of identical 1D barriers. M is the transfer matrix for the barrier and is the same for all. P is the propagation matrix over free space and differs due to randomness in the system.}
    \label{fig:3}
\end{figure}

We now consider an infinite one-dimensional (1D) array of identical  barriers separated by a mean distance $d$ (Fig. \ref{fig:3}). The distances between the barriers is random, i.e., the distance between two consecutive barriers is $d+{\delta}$ where ${\delta} (\ll d)$ is random. We will now study localization in this 1D array due to presence of the randomness in position of the barriers.

As the barriers are identical, their transfer matrices are equal, given by (13). The free propagation matrix differs between each barrier because of randomness. It retains the same form as (12) but with  $p$ in place of $q$ as the propagation is now over free space. The presence of randomness in the array is reflected in the propagation matrix in the form of the distance between two consecutive barriers: $d+{\delta}$.

\begin{figure}[ht]
    \includegraphics[width=.75\textwidth]{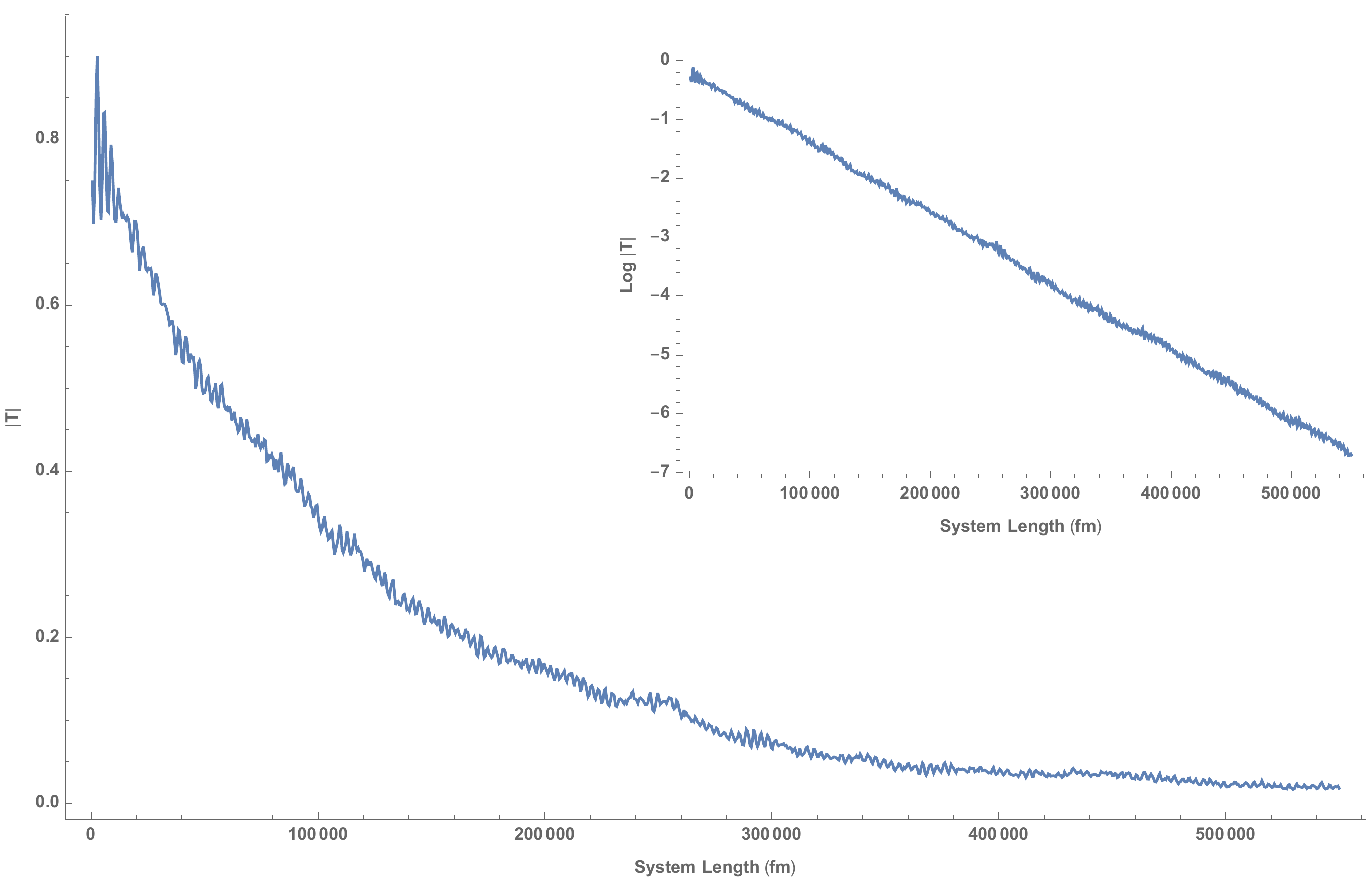}
    \caption{Exponential localization of the wave function of an electron. Here, the barrier height and width were chosen to be 10 MeV and 400 fm respectively, mean barrier distance d as 150 fm, incident particle momentum as 11 MeV/c and disorder strength W as 2. The plot above is the averaged result of 100 iterations.}
    \label{fig:4}
\end{figure}

The transfer matrix approach reduces the problem of infinite barriers to one involving product of random matrices. As transfer matrices follow the composition law \cite{Markos}, the total transfer matrix M of a system of $N$ barriers can be written as :

\begin{alignat}{1}
M_{Total} = M P_{N-1} M . . . P_2 M P_1 M
\end{alignat}
where $M$ is given by (13) and $P_i$ has the form:
\begin{equation}
P_{i} = 
\begin{pmatrix}
		e^{\frac{ip(d+\delta_i)}{\hbar}} & 0\\
        0 & e^{\frac{-ip(d+\delta_i)}{\hbar}} \\
\end{pmatrix}.
\end{equation}\\
The first diagonal element represents a particle traveling towards the right and the second a particle traveling towards the left. The matrix $M_{Total}$ is the transfer matrix for an array of $N$ identical barriers with disorder in position. It is easy to see that $M_{Total}$ is given by the product of random matrices and for a large value of $N$ it would model the situation in Fig. \ref{fig:4} suitably.\\

We began by placing a constraint on the energy of the incident particle $E > V_0$. With this it is expected that for $E \gg V_0$, the localization length tends to infinity. However, the same cannot be said for energies much closer to $V_0$.

The product given in (20) was computed using \textit{MATHEMATICA} for 1000 identical barriers and the transmission coefficient was calculated using (18). The term $\delta$ was generated randomly from a normal distribution with mean zero and standard deviation $W$.  The parameters were given in nuclear units with the barrier height and width as 10 MeV and 400 fm respectively, mean barrier distance d as 150 fm, incident particle momentum as 11 MeV/c and disorder strength W as 2. The plot shown in Figure 5 is the averaged result of 100 iterations.\\

As the number of barriers increase, the transmission coefficient tends to zero exponentially. This exponential localization is also confirmed by an exponential curve fitted to the data. The strength of this localization is dependent on the strength of the disorder, namely $W$ for weak disorder. As we increase the disorder strength, we attain a constant value.

In Fig. \ref{fig:4}, the transmission probability was calculated after every successive barrier and averaged over 100 iterations. This transmission probability was then plotted alongside the system length. As we are varying the inter-barrier distances to introduce randomness into the system, they subsequently had minor differences in each iteration. Hence, to plot the transmission probability calculated after every barrier with system length, the inter-barrier distances were averaged as well.

Essentially, this means that both quantities on the X and Y axis are averaged results for a large number of iterations (100 in this case).\\

We can further discuss the Lyapunov exponent $\gamma$ and the localization length $\xi$ for the above system. The two are related and defined as:
\begin{equation}
\gamma = \frac{1}{\xi} = -\lim\limits_{L \to \infty}\frac{ < log_e |T| >}{L}
\end{equation}
where L is the total length of the chain $L=N*(a+d)$.\\

\begin{figure}[ht]
    \includegraphics[width=.49\textwidth]{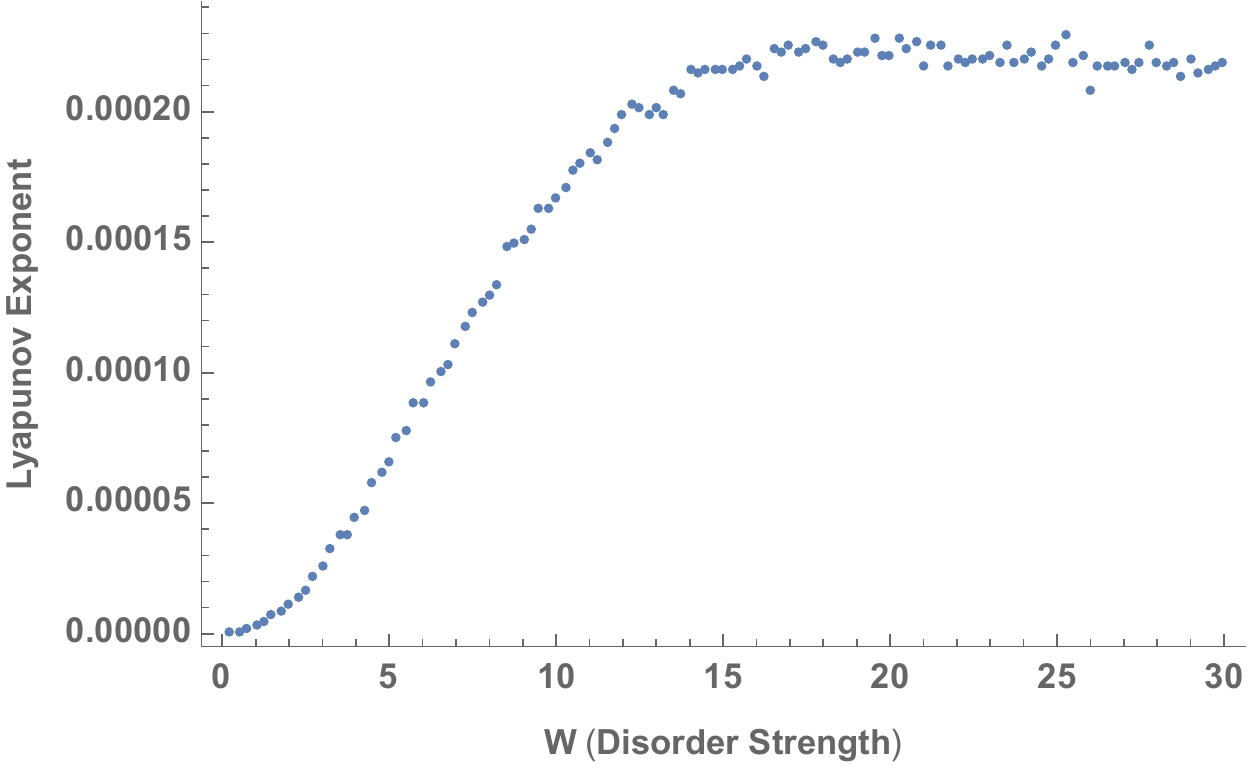}
    \includegraphics[width=.49\textwidth]{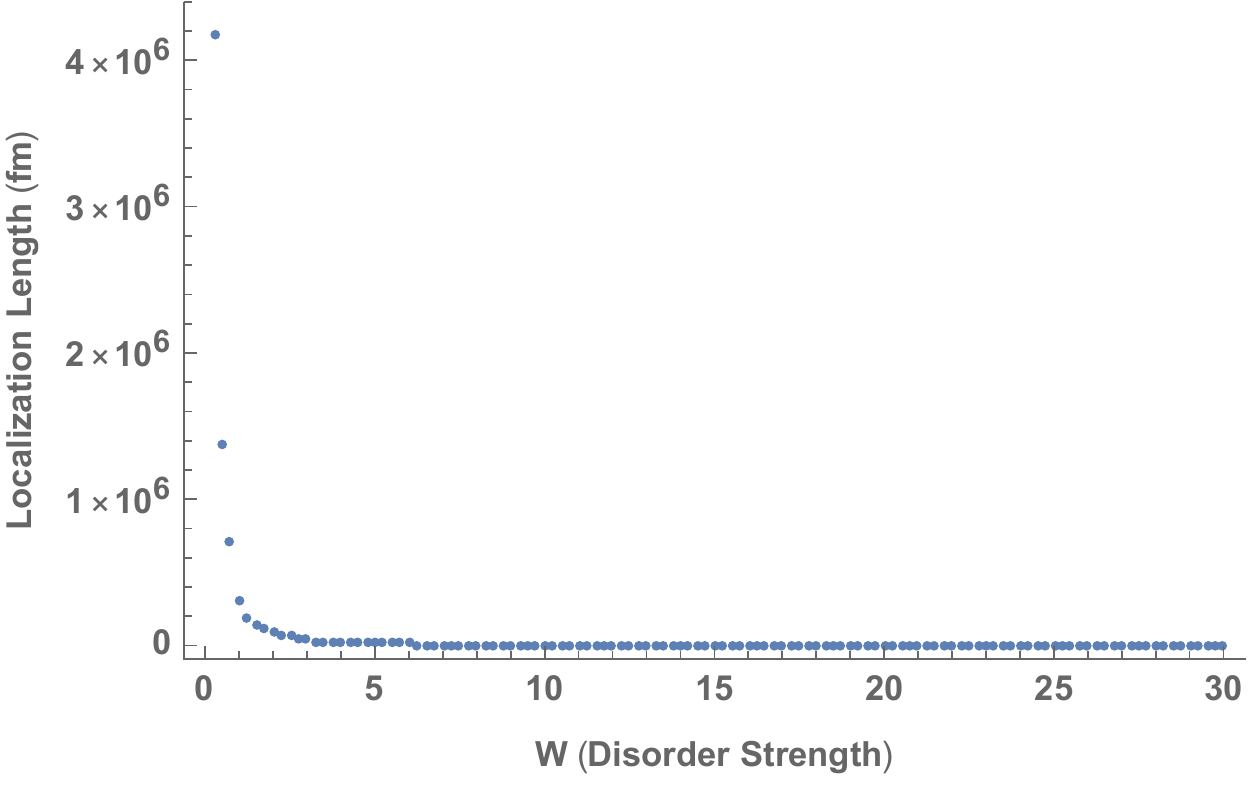}
    \caption{The Lyapunov exponent (left) and the localisation length (right) as a function of disorder strength for an electron. The incident energy of the electron here is 11 MeV/c and the system parameters are the same as in Fig 4. Each point on the two plots here is the calculated result averaged over 50 iterations.  }
    \label{fig:5}
\end{figure}

We now look for a relation between the localization length and the disorder strength of the system. We expect that the relation between the two quantities can be expressed as a power law. This can be confirmed if we plot a Log-Log plot of the two quantities and obtain a straight line. This straight line would then correspond to a power law relation which would give us-
\begin{alignat}{1}
    \xi &\propto W^c \\
    \text{Log}[\xi] &\propto c \text{ Log}[w]
\end{alignat}
\begin{figure}[ht]
    \includegraphics[width=0.6\textwidth]{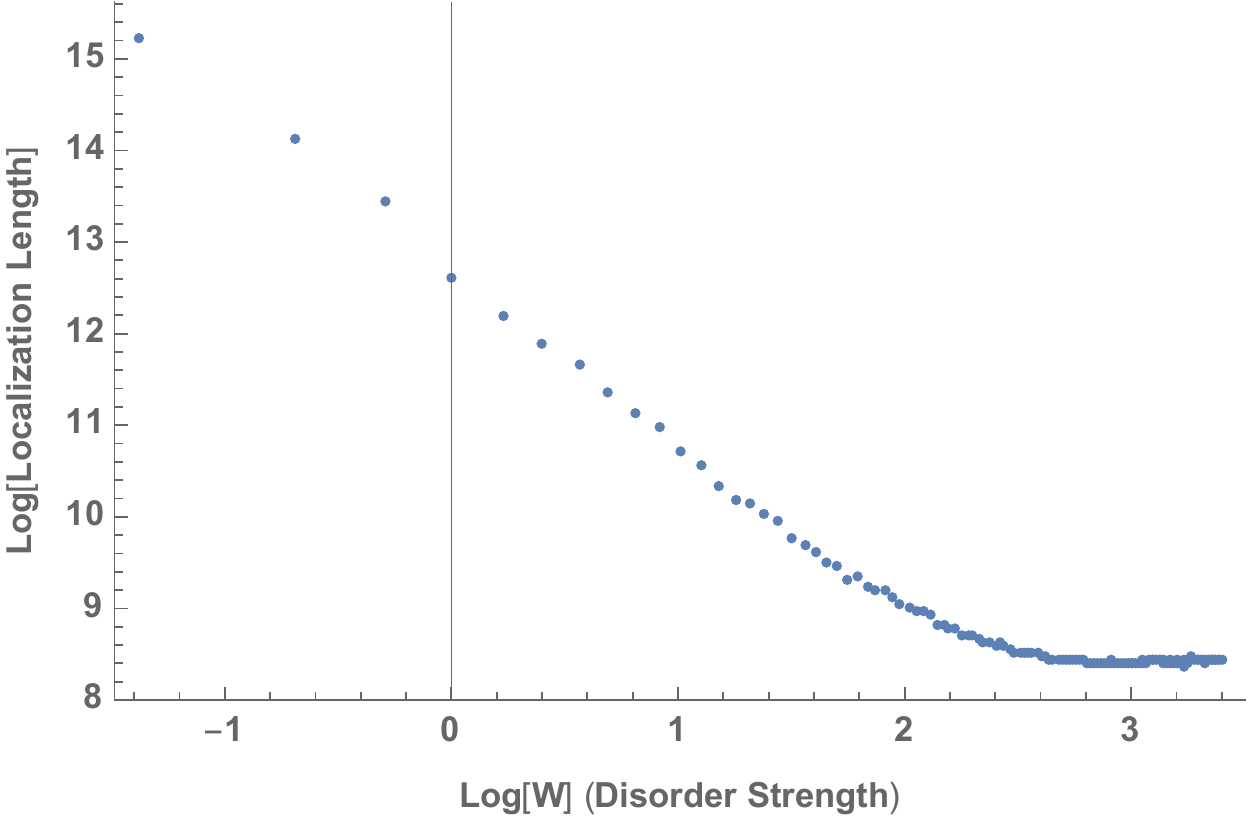}
    \caption{The Log-Log plot of the localisation length vs disorder strength plot in Fig. \ref{fig:5} (right). A straight line in the Log-Log plot corresponds to a power law relation between the two quantities. Here, we see that for small disorder strengths there is a linear relationship which does not hold for larger values of the same.}
    \label{fig:6}
\end{figure}

In a Log-Log plot as described above (Fig. \ref{fig:6}), we see a linear relationship when the disorder strength is small, which is broken at large disorder strengths. The region where the slope of the graph approaches zero is where the disorder strength is too large and the approximation of small perturbations in the inter-barrier distance no longer holds.\\

\begin{figure}[ht]
    \includegraphics[width=.6\textwidth]{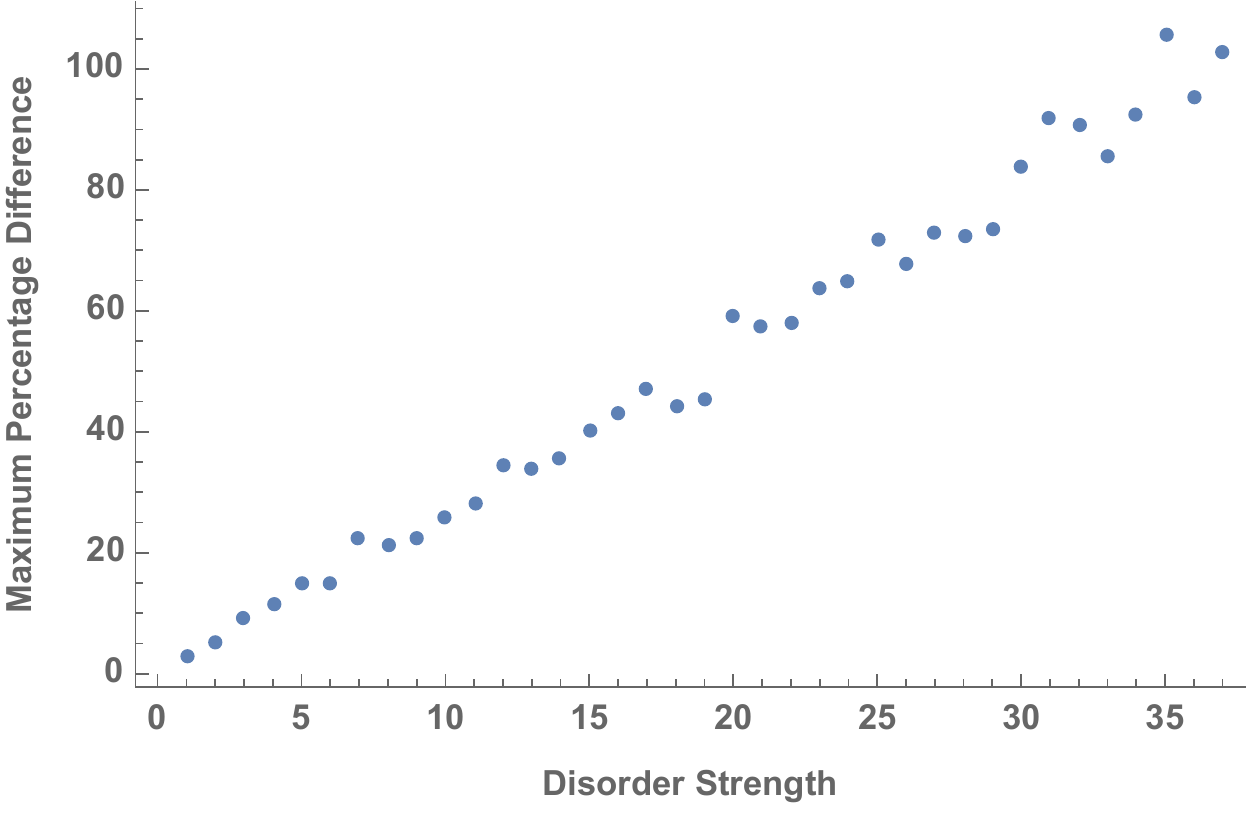}
    \caption{The plot shows the maximum percentage difference between ideal mean inter-barrier distance and the perturbed value in the presence of disorder in relation to the disorder in the system. The ideal inter-barrier distance corresponds to 150 fm (d) here and the perturbed value is (d+$\delta$) where $\delta$ is generated randomly in each iteration.  For high disorder strengths, we see that the is difference is very large and cannot be classified as ``small" perturbations.}
    \label{fig:7}
\end{figure}

To verify this reasoning, we plot (Fig. \ref{fig:7}) the maximum percentage difference between the ideal mean inter barrier distance and the new value obtained after introducing disorder. We see that even for W=10, the maximum difference is upto 20\% for a mean inter-barrier distance of 150 fm. For even higher values of disorder strength, the condition of small perturbations breaks completely.

\begin{figure}[ht]
    \includegraphics[width=\textwidth]{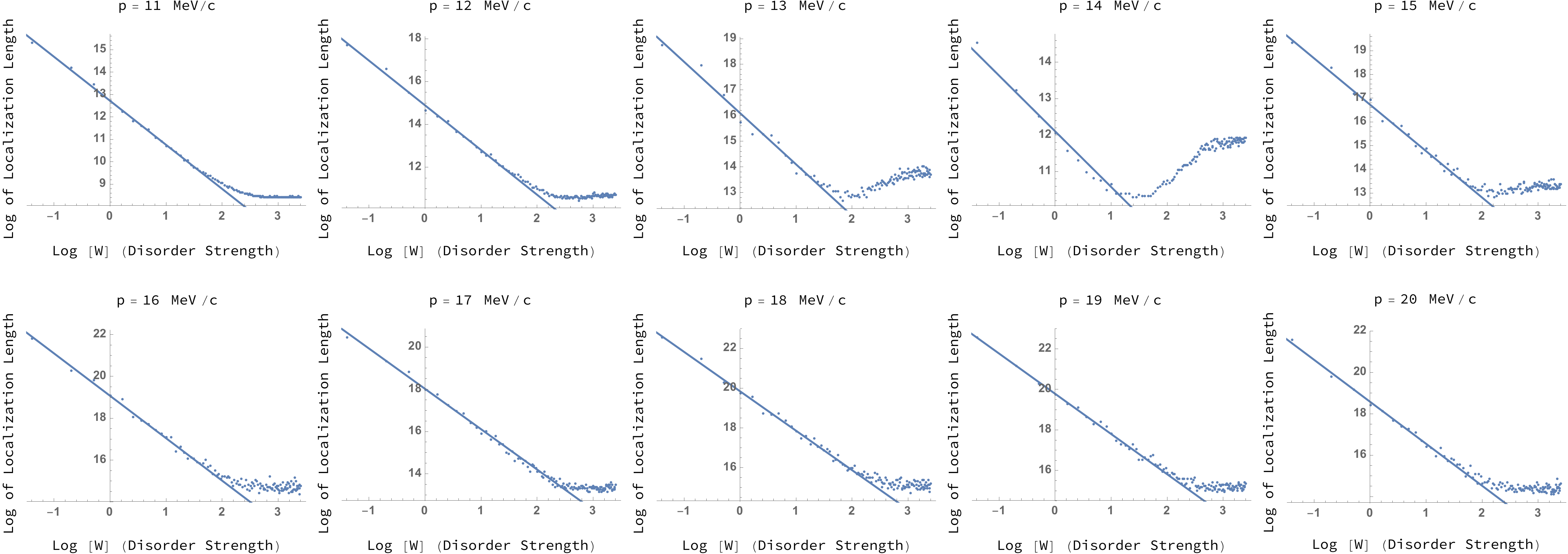}
    \caption{The Log-Log plots of the localisation length vs disorder strength plot for different values of incident momentum of an electron. The height of the barrier is 10 MeV in all cases and the remaining system parameters are the same as described in Fig. \ref{fig:4}. The straight line fit in each case fits the first 16 points in the linear region going from a disorder strength of 0.25 to 4.}
    \label{fig:8}
\end{figure}

\newpage
In order to examine any variation in the linear behaviour at small disorder strengths with incident energy of the particle (electron), we plot the Log-Log plots (Fig. \ref{fig:8}) for different values of incident momentum.\\

The slope (c) for successive values of incident momentum was found to be as:\\
-1.96253 (11 MeV/c), -2.08447 (12 MeV/c), -2.00154 (13 MeV/c), -1.5271 (14 MeV/c), -1.95633 (15 MeV/c), -2.02576 (16 MeV/c), -1.90843 (17 MeV/c),-1.98469 (18 MeV/c), -1.98425 (19 MeV/c), -2.02876 (20 MeV/c).\\

An average value of the constant $c$ in (23) was calculated to be $-1.95\pm0.15$.

\newpage
\section{Three component momentum}
\subsection{Transfer matrix in presence of spin flip}
\begin{figure}[ht]
    \centering
    \includegraphics[width=0.75\textwidth]{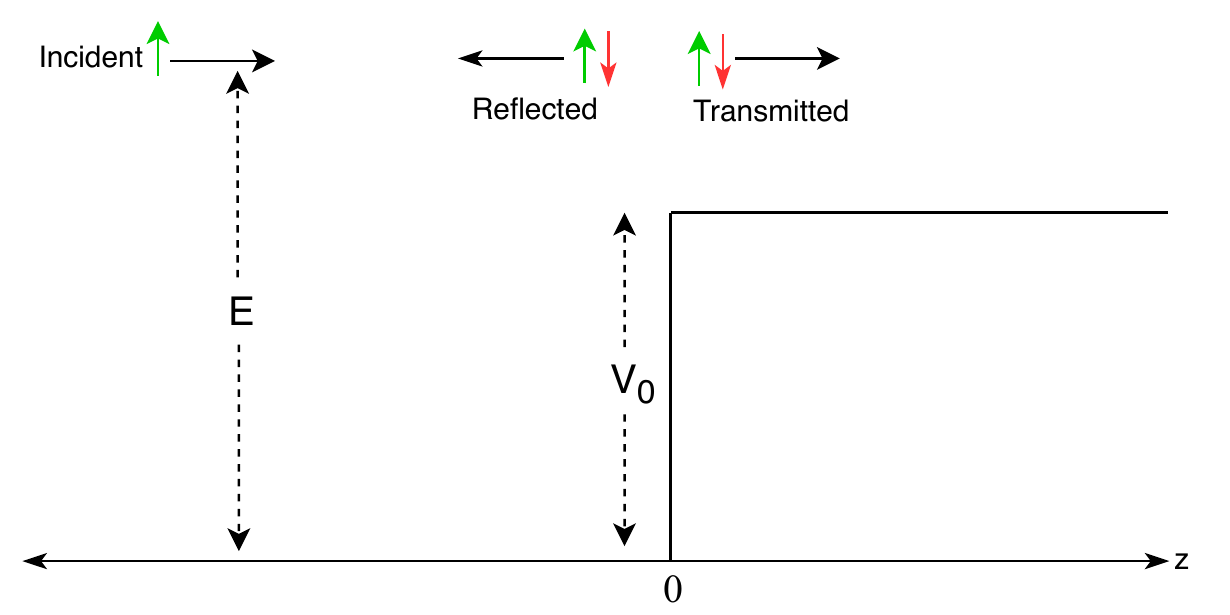}
    \caption{Pictorial representation of spin-flip of an incident Dirac particle in 1D as described in \cite{Glass}. The green arrow represents spin  in the 'up' direction and the red arrow represents spin in the 'down' direction.}
    \label{fig:9}
\end{figure}

\begin{figure}[ht]
    \centering
    \includegraphics[width=0.75\textwidth]{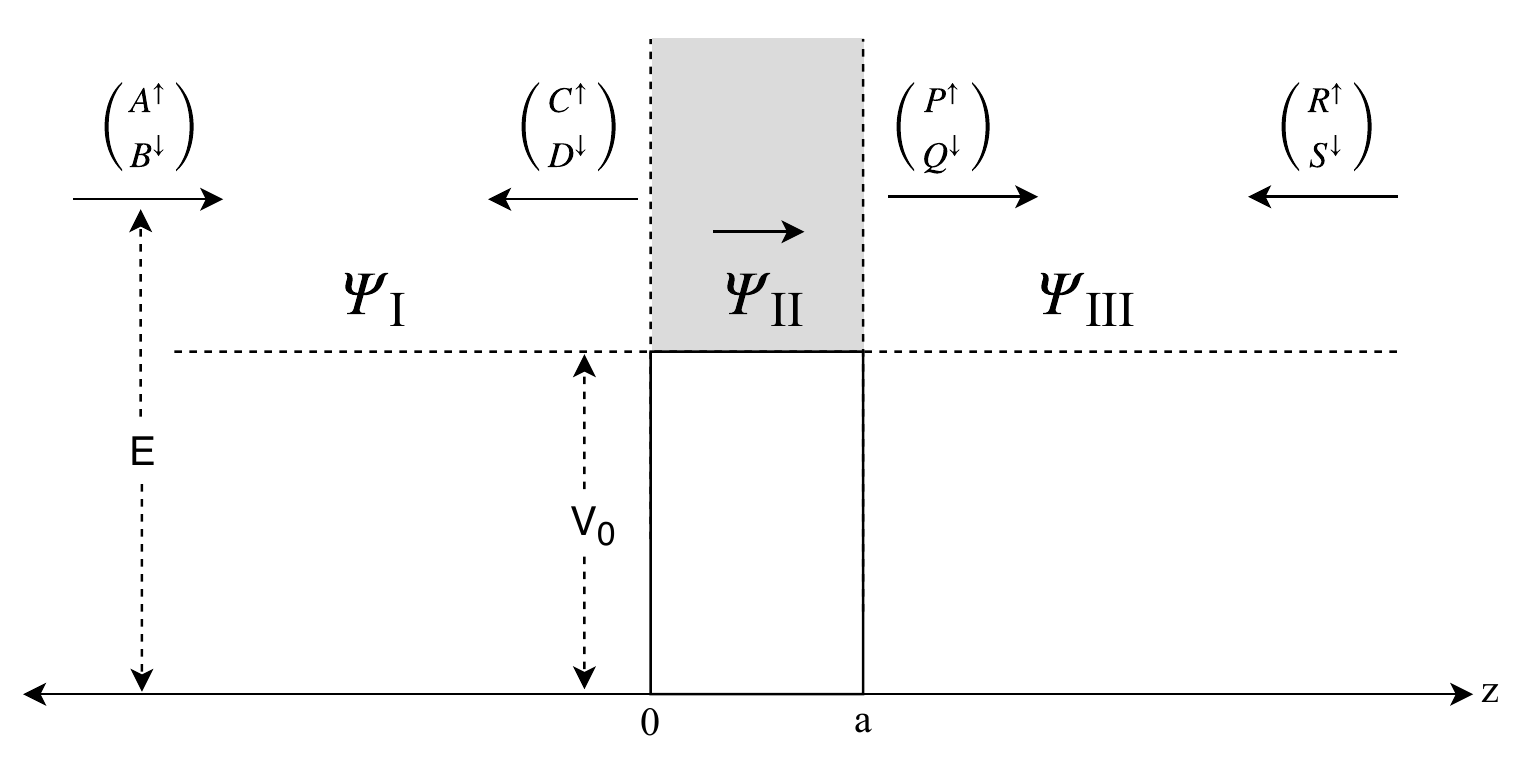}
    \caption{Pictorial representation of a Dirac particle incident on a potential barrier with $E>V_{0}$ with spin-flip incorporated. The particle here posses transverse momentum as well. $A$ and $B$ are the incident spin-up and spin-down amplitudes from the left respectively, while $C$ and $D$ are the reflected spin-up and spin-down amplitudes respectively. Similarly, $P$ and $Q$ are the transmitted spin-up and spin-down amplitudes moving towards the right and $R$ and $S$ are the spin-up and spin-down incident amplitudes for a particle incident from the right. Amplitudes in region II are present but not marked for clarity.}
    \label{fig:10}
\end{figure}

In certain situations, a  1D Dirac scattering over a potential barrier gives rise to spin-flips in transmission of the particles \cite{Glass}. The spin-flips are a result of the  transverse momentum possessed by the particle and disappear when these  components vanish. This is visualized in Fig.  \ref{fig:9} and Fig. \ref{fig:10}. We extend this for a rectangular barrier, and subsequently to a disordered chain of barriers.\\

The general positive energy solutions of the Dirac equation can again be written as (4). In this case however, we cannot reduce our $4 \times 4$ matrix to a $2 \times 2$ matrix due to the presence of the transverse momentum $p_{x}$ and $p_{y}$.\\

We follow our previous approach to work out the transfer matrix for this case. 
\begin{equation}
\begin{pmatrix}
		P^{\uparrow}  e^{\frac{ip_{z}a}{\hbar}} \\
        Q^{\downarrow}  e^{\frac{ip_{z}a}{\hbar}} \\
        R^{\uparrow}  e^{\frac{-ip_{z}a}{\hbar}} \\
        S^{\downarrow}  e^{\frac{-ip_{z}a}{\hbar}} \\
\end{pmatrix}  = M
\begin{pmatrix}
		A^{\uparrow} \\
        B^{\downarrow} \\
        C^{\uparrow} \\
        D^{\downarrow} \\
\end{pmatrix}.
\end{equation}
Here $A$ and $B$ are the incident spin-up and spin-down amplitudes from the left respectively, while $C$ and $D$ are the reflected spin-up and spin-down amplitudes respectively. Similarly, $P$ and $Q$ are the transmitted spin-up and spin-down amplitudes moving towards the right and $R$ and $S$ are the spin-up and spin-down incident amplitudes for a particle incident from the right (if any).\\

To simplify we find the transfer matrix $M_{step}$ for the potential step at $z = 0$ satisfying the equivalent relation (7) for the spin-flip case-
\begin{equation}
\begin{pmatrix}
		E^{\uparrow} \\
        F^{\downarrow} \\
        G^{\uparrow} \\
        H^{\downarrow} \\
\end{pmatrix}  = M_{step}
\begin{pmatrix}
		A^{\uparrow}\\
        B^{\downarrow} \\
        C^{\uparrow} \\
        D^{\downarrow} \\
\end{pmatrix}.
\end{equation}
Where $E$, $F$, $G$ and $H$ are amplitudes of the wavefunction within the barrier region.
Note that by conservation of momentum $p_{x} = q_{x}$ and $p_{y} = q_{y}$. Evaluating the spinors, we obtain the matrix equation :
\begin{equation}
\begin{pmatrix}
		1 & 0 & 1 & 0 \\
        0 & 1 & 0 & 1\\
        p_{z} & p_{-} & -p_{z} & p_{-} \\
        p_{+} & -p_{z} & p_{+} & p_{z}
\end{pmatrix} 
\begin{pmatrix}
		A^{\uparrow}\\
        B^{\downarrow} \\
        C^{\uparrow} \\
        D^{\downarrow} \\
\end{pmatrix}  =
\begin{pmatrix}
		1 & 0 & 1 & 0 \\
        0 & 1 & 0 & 1\\
        rq_{z} & rp_{-} & -rq_{z} & rp_{-} \\
        rp_{+} & -rq_{z} & rp_{+} & rq_{z}
\end{pmatrix} 
\begin{pmatrix}
		E^{\uparrow} \\
        F^{\downarrow} \\
        G^{\uparrow} \\
        H^{\downarrow} \\
\end{pmatrix},
\end{equation}
where 
\begin{equation}
p_{\pm} = p_{x} \pm i p_{y}. 
\end{equation}
$M_{step}$ is then given by
\begin{equation}
 M_{step} =
\begin{pmatrix}
 	\frac{1}{2} \left(\frac{p}{q_z r}+1\right) & -\frac{ p_- (r-1)}{2 q_z r} & \frac{1}{2}-\frac{p}{2 q_z r} & -\frac{p_- (r-1)}{2 q_z r} \\
	 \frac{p_+ (r-1)}{2 q_z r} & \frac{1}{2} \left(\frac{p_z}{q_z r}+1\right) & \frac{p_+ (r-1)}{2 q_z r} & \frac{1}{2}-\frac{p}{2 q_z r} \\
	 \frac{1}{2}-\frac{p_z}{2 q_z r} & \frac{p_- (r-1)}{2 q_z r} & \frac{1}{2} \left(\frac{p_z}{q_z r}+1\right) & \frac{p_- (r-1)}{2 q_z r} \\
	 -\frac{p_+ (r-1)}{2 q_z r} & \frac{1}{2}-\frac{p_z}{2 q_z r} & -\frac{p_+ (r-1)}{2 q_z r} & \frac{1}{2} \left(\frac{p_z}{q_z r}+1\right) \\

\end{pmatrix}.
\end{equation}

The transfer matrix $M_{step}$ in (29) reduces to the one in (10) in the absence of spin flip ($p_x = p_y = 0$). The transfer matrix $M$ for the rectangular potential in the presence of spin flip can now be worked out with (11) as before. The free propagation transfer matrix $M_{0}$, however, is now a $4 \times 4$ matrix.\\
\begin{equation}
M_{0} = 
\begin{pmatrix}
		e^{\frac{iq_{z}a}{\hbar}} & 0 & 0 & 0\\
		0 & e^{\frac{iq_{z}a}{\hbar}} & 0 & 0\\
        0 & 0 & e^{\frac{-iq_{z}a}{\hbar}} & 0\\
        0 & 0 & 0 & e^{\frac{-iq_{z}a}{\hbar}}\\
\end{pmatrix}
\end{equation}
\\
The transfer matrix $M$ :
\begin{equation}
M = 
\begin{pmatrix}
		u & 0 & v^* & w^*\\
        
        0 & u & w & v^*\\
        
        v & w^* & u^*& 0 \\
        
        w & v & 0 & u^*
        
\end{pmatrix},
\end{equation}
with $u$, $v$, $w$ as
\begin{alignat}{1}
u &= \cos \left(\frac{aq_{z}}{\hbar}\right) + i\alpha_{+}\sin \left(\frac{aq_{z}}{\hbar}\right),\nonumber \\
v &= +i\alpha_{-}\sin \left(\frac{aq_{z}}{\hbar}\right),\\
w &= i\mu_{\pm}\sin \left(\frac{aq_{z}}{\hbar}\right).\nonumber
\end{alignat}
Here,
\begin{alignat}{1}
\alpha_{\pm} &= \frac{1}{2}\left[ \left( \frac{p_{z}}{rq_{z}} \pm \frac{rq_{z}}{p_{z}}\right) \pm \frac{(p_{x}^2 + p_{y}^2)(r-1)^2}{p_{z}q_{z}r}\right] ,\\
\mu_{\pm} &= \frac{p_{\pm}(r-1)}{rq_{z}}, \nonumber 
\end{alignat}
and $p_{\pm}$ are given by (28). Note that $\mu_{-}$ is conjugate to $\mu_{+}$. The matrix element $w$ disappears in the absence of transverse momentum components and $\alpha_{\pm}$ reduce back to (15). The $4 \times 4$ transfer matrix in (28) reduces to the $2 \times 2$ transfer matrix in (13) in the absence of spin-flip.\\

Like the previous case, the transfer matrix $M$ maintains both time reversal symmetry and conservation of current density. This is again verified \cite{Markos} by the calculation of the following relations:
\begin{alignat}{1}
\mbox{Det } &M = 1 \quad \mbox{for time-reversal symmetry}\\
M^\dagger 
\begin{pmatrix}
		I_2 & 0\\
        0 & -I_2 \\
\end{pmatrix} M &= \begin{pmatrix}
		I_2 & 0\\
        0 & -I_2 \\
\end{pmatrix} \quad \mbox{for conservation of current density}
\end{alignat}
where $I_2$ is the $2\times2$ identity matrix.\\

From the transfer matrix (31), we see that the transmission of a polarized incident particle remains spin-up polarized. This is unlike transmission over a potential step \cite{Glass}, which had a spin-flip components in transmission. \\

The transmission coefficient $T_z$ (transmission in the z direction) is now given by the matrix elements $M_{33}$ and $M_{44}$. These elements correspond to the transmission of spin-up (down) components when the incident particle is spin-up (down). Since these are independent of each other, for a unit incidence they are equal. 
\begin{equation}
    T_z = |t|^2 = \frac{1}{|M_{33}|^2} = \frac{1}{|M_{44}|^2} 
\end{equation}
Using the expression for $M_{33}$ from (31) we obtain $T_z$ as:
\begin{equation}
    T_z = \bigg[1 + \frac{1}{4}\left\{ \left( \frac{p_z}{rq_z} - \frac{rq_z}{p_z} \right)^2 + \frac{2\eta(p_z^2 +rq_z^2) + \eta^2}{p_z^2 q_z^2 r^2}\right\} \sin ^2 \bigg[\frac{aq_z}{\hbar}\bigg]\bigg]^{-1}
\end{equation}
where $\eta= (p_x^2 + p_y^2)(r-1)^2 = p_+ p_- (r-1)^2 $. Note that in absence of spin-flip, $\eta$ becomes zero and (37) reduces to (19). The presence of $\eta$ here also indicates that the transmission coefficient is lower when spin-flips are possible. Thus, we would expect localization to occur at much shorter lengths as compared to the earlier case.

\subsection{Localization in one dimension}
Our approach here follows that of Section 3.2. The transfer matrix for the barrier is now given by (31) and randomness in the system is reflected in the propagation matrix. The propagation matrix is now a $4\times4$ matrix of the form,
\begin{equation}
    P_{i} = 
    \begin{pmatrix}
		e^{\frac{ip(d+\delta_i)}{\hbar}} & 0 & 0 & 0 \\
        0 & e^{\frac{ip(d+\delta_i)}{\hbar}} & 0 & 0 \\
        0 & 0 & e^{\frac{-ip(d+\delta_i)}{\hbar}} & 0 \\
        0 & 0 & 0 & e^{\frac{-ip(d+\delta_i)}{\hbar}} \\
    \end{pmatrix}.
\end{equation}
The first and second diagonal elements represent a spin-up and spin-down particle respectively, traveling towards the right. Similarly the last two represent a spin-up and spin-down particle traveling towards the left. The transfer matrix $M_{total}$ for an array of $N$ identical barriers is given by (20). As mentioned before, the presence of the additional  term involving $\eta$ in the transmission probability greatly affects the transmission of the particle.\\

Our constraints on the system are the same as in Section 3.3. The product given in (20) was computed employing \textit{MATHEMATICA} for 500 identical barriers and the transmission coefficient was calculated using (37). Here, the barrier height and width were chosen to be 10 MeV and 400 fm respectively, mean barrier distance d as 150 fm and disorder strength W as 2 for both the spin-flip and no spin-flip case. An important difference here is to specify the incident angle of the particle on the barrier as the incident momentum is three dimensional. For this case, the particle is chosen to be incident at an angle of $45^{\circ}$ with respect to the z-axis.

\begin{figure}[ht]
    \includegraphics[width=.47\textwidth]{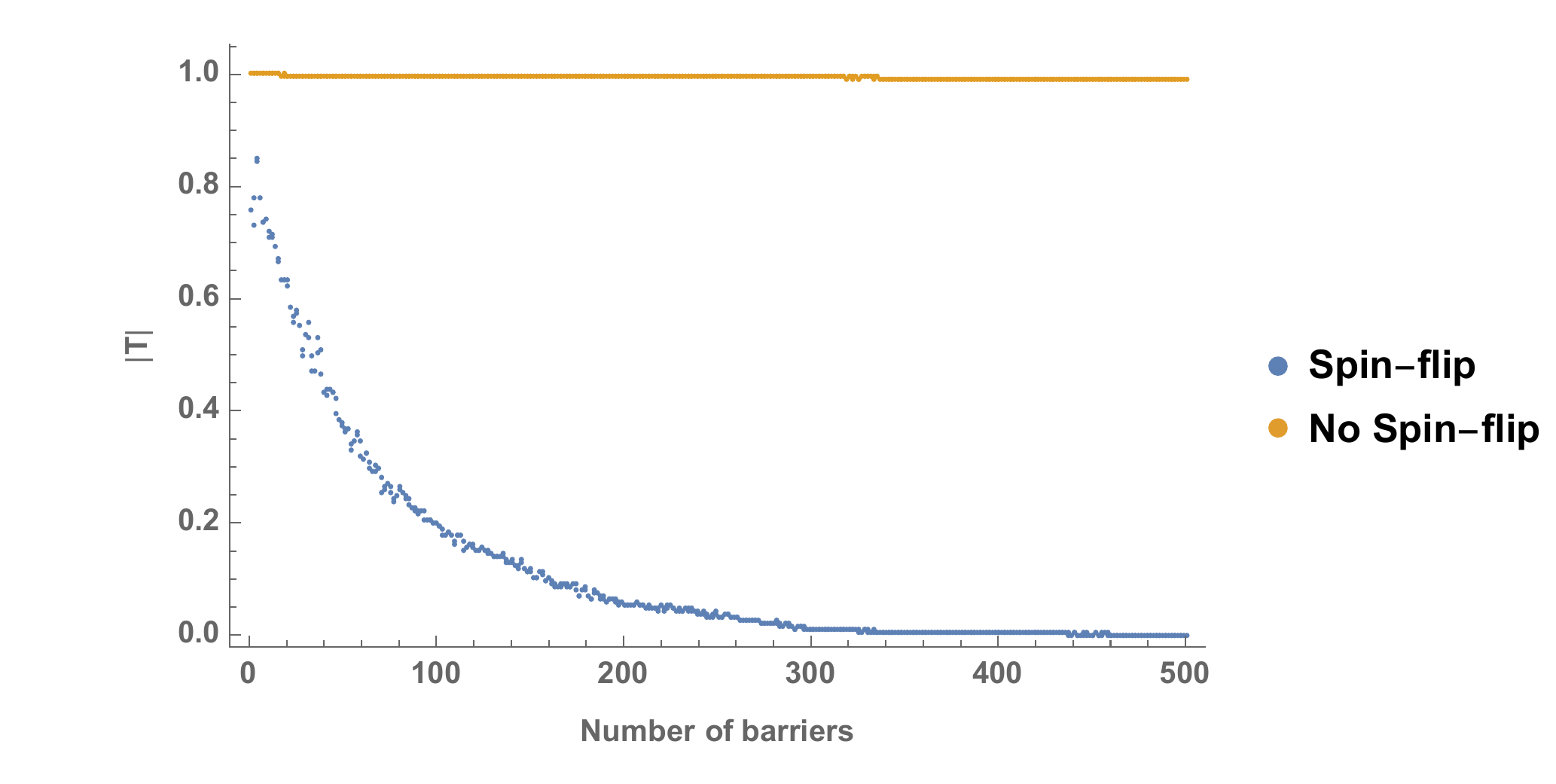}
    \includegraphics[width=.47\textwidth]{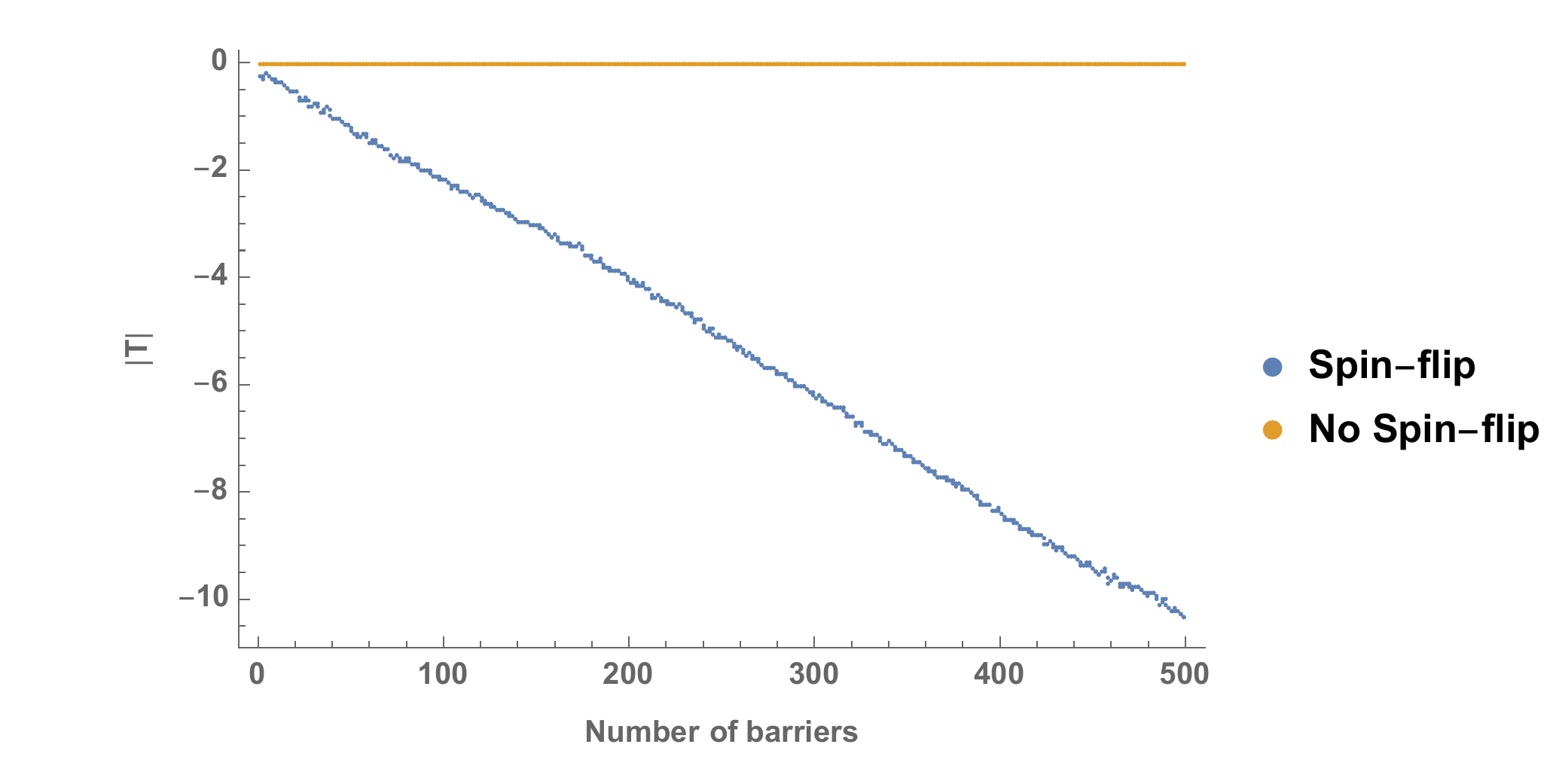}
    \caption{Exponential localization of the wave function of electron when possessing transverse momentum components. Here, the barrier height and width were chosen to be 10 MeV and 400 fm respectively, mean barrier distance d as 150 fm and disorder strength W as 2 for both the spin-flip and no spin-flip case. For the spin-flip case, the particle was incident on the x-z plane at an angle $45^{\circ}$ with a total energy of 28.2889 MeV and 20.0065 MeV for the no-spin flip particle possessing momentum only in the z direction. These energies were taken such that the incident momentum of both the particles in the z direction were equal to 20 MeV/c. The plot above is the averaged result of 100 iterations}
    \label{fig:11}
\end{figure}

Fig. \ref{fig:11} shows the presence of an exponential localization of the wave function in the $z$ direction for the spin-flip case. As the number of barriers increase, the transmission coefficient tends to zero exponentially. This exponential localization is also confirmed by an exponential curve-fit on the data. 

Fig. \ref{fig:11} also shows a key distinction between the two cases of spin-flip and no spin-flip. The localization is much faster in the case of spin-flip because of the additional $\eta$ term in the denominator. However, in this case the localization is only along the z direction, the x and y components are unaffected.

\section{Conclusions}
We have established localisation in 1D relativistic systems for a spin $\frac{1}{2}$ particle through a mix of theoretical and numerical methods. We  considered two different relativistic systems. The first ``ordinary" case where the particle is moving in 1D and is incident on a 1D barrier and the second where the particle is moving in 3D yet is incident on a barrier that is 1D. We see that localisation in the second case is much ``quicker" than in the first case. We also explored the Lyapunov exponent and the localisation length of an electron for the former case. On the basis of a numerical computation, we also determined the dependence of localisation length on the disorder strength arriving at a relation of $\xi \propto W^{(-1.95\pm0.15)}$.\\

In the realm of future studies, we see many openings following a similar line of work. An interesting example would be the extension of this work into the energy regime of $E+mc^2<V$ where-in the effect of the Klein paradox has to be considered. Pair-production in such a scenario could lead to unexpected results. Furthermore, an exhaustive analysis of the spin-flip case for localisation length and density of states is also a very exciting prospect. 

\section{Author contribution statement}

All the authors were equally involved in the work, from conception to conclusions.

\end{document}